\documentclass[showpacs,amsmath,amssymb,twocolumn]{revtex4}
\usepackage{graphicx}
\usepackage{dcolumn}
\usepackage{bm}

\begin{document}

\title{\bf\large{LOFF Pairing vs. Breached Pairing in Asymmetric Fermion Superfluids}}
\author{\normalsize{Lianyi He\footnote{Email address: hely04@mails.tsinghua.edu.cn}, Meng Jin\footnote{Email address: jin-m@mail.tsinghua.edu.cn
        } and Pengfei Zhuang\footnote{Email address:
        zhuangpf@mail.tsinghua.edu.cn}}}
\affiliation{Physics Department, Tsinghua University, Beijing
100084, China}

\begin{abstract}
A general analysis for the competition between breached pairing
(BP) and LOFF pairing mechanisms in asymmetric fermion superfluids
is presented in the frame of a four fermion interaction model. Two
physical conditions which can induce mismatched Fermi surfaces are
considered: (1) fixed chemical potential asymmetry $\delta\mu$ and
(2) fixed fermion number asymmetry $\alpha$. In case (1), the BP
state is ruled out because of Sarma instability, and the LOFF
state is thermodynamically stable in a narrow window of
$\delta\mu$. In case (2), while the Sarma instability can be
avoided and both the BP and LOFF states can survive provided
$\alpha$ is less than the corresponding critical value, the BP
state suffers magnetic instability and the LOFF state is always
thermodynamically stable. While the LOFF window in case (2) is
much larger than the one in the conventional case (1), the
longitudinal superfluid density of the LOFF state is negative for
small $\alpha$ and it suffers magnetic instability too.
\end{abstract}


\maketitle

\section {Introduction}
\label{s1}
The fermion pairing between different species with mismatched
Fermi surfaces, which was discussed many years ago, promoted new
interest in both theoretic and experimental studies in recent
years. The mismatched Fermi surfaces can be realized, for
instance, in a superconductor in an external magnetic
field\cite{sarma} or a strong spin-exchange
field\cite{larkin,fulde,takada}, an electronic gas with two
species of electrons from different bands\cite{liu}, a
superconductor with overlapping bands\cite{suhl,kondo}, trapped
ions with dipolar interactions\cite{cirac}, a mixture of two types
of fermionic cold atoms with different densities and/or
masses\cite{liu,bedaque,caldas,exp1,exp2}, an isospin asymmetric
nuclear matter with proton-neutron pairing\cite{sedrakian}, and a
neutral dense quark
matter\cite{schafer,alford,rajagopal,steiner,alford2,huang,shovkovy,huang2,alford3,abuki,ruster,blaschke,lawley}.
In the early studies of superconductivity, Sarma\cite{sarma} found
a spatially isotropic and uniform state where there exist gapless
modes, namely the excitation of these quasiparticle modes needs no
energy. However, this state is energetically unstable, compared
with the fully gapped BCS state. This is the so-called Sarma
instability\cite{sarma}. Recently, this instability was widely
discussed in general case\cite{forbes,liao} and in neutral color
superconductivity\cite{shovkovy,huang2}. It is now widely accepted
that the Sarma instability can be avoided through two possible
ways, fixed number difference of the two kinds of
fermions\cite{shovkovy,huang2,forbes,liao} or a proper momentum
structure of the attractive interaction between the two
species\cite{forbes}. The Sarma state is now also called breached
pairing (BP) state, since in this state the dispersion relation of
one branch of the quasi-particles has two zero points at momenta
$p_1$ and $p_2$ where gapless excitations happen, and the
superfluid component in the regions $p<p_1$ and $p>p_2$ is
breached by a normal component in the region $p_1<p<p_2$.

The LOFF state which is a spatially anisotropic ground state was
proposed in the study of superconductors in a strong spin-exchange
field by Larkin, Ovchinnikov, Fulde and
Ferrell\cite{larkin,fulde}. In this ground state, the rotational
symmetry and/or the translational symmetry of the system are
spontaneously broken. In the study of asymmetric nuclear matter,
asymmetric atomic fermion gas and color superconductivity, the
LOFF phase has been widely investigated
\cite{alford5,bowers,leibovich,casalbuoni3,casalbuoni4,casalbuoni5,sedrakian2,muether,machida,sedrakian3,castorina,mizhushima,yang,son,sheehy,yang2,dukelsky}.

While the Sarma instability can be avoided, a new dynamical
instability of the BP state which is called magnetic instability
in the literatures arises in the study of dense quark matter. It
is found that the response to external color magnetic field in
neutral dense quark matter in the BP state is paramagnetic, i.e.,
the squared Meissner masses of some gluons become
negative\cite{huang3,huang4,casalbuoni,alford4,fukushima}. This
instability was also studied in atomic fermion system\cite{wu}. A
general analysis for asymmetric superconductors shows that this
dynamical instability is related to the breached pairing mechanism
only and independent of the details of the attractive
interactions\cite{he}. It is recently pointed out that, due to the
magnetic instability, the free energy of the LOFF state or some
other exotic anisotropic states is probably lower than that of the
BP
state\cite{he,giannakis,giannakis2,giannakis3,huang5,hong,gorbar,gorbar2,casalbuoni2}.
However, this conclusion is obtained from the study for asymmetric
systems with fixed chemical potentials. In many physical cases we
are interested in, the chemical potentials are not directly fixed,
instead the particle numbers are fixed or some other constraints
on the particle numbers are required. For these systems, the
magnetic instability is still not clear. In this paper we will
give a general and systematic comparison between the BP and LOFF
states in asymmetric fermion superfluids under different
conditions which can lead to mismatched Fermi surfaces.

The paper is organized as follows. In Section \ref{s2} we briefly
review the formalism for the isotropic BCS, BP and anisotropic
LOFF states. In Section \ref{s3}, we analytically discuss the
thermodynamic and dynamical instabilities of the BP state, and
show why the LOFF state may be more favored than the BP state. In
Sections \ref{s4} and \ref{s5}, we discuss, respectively, the
systems with fixed chemical potential asymmetry and fixed density
asymmetry. We summarize in Section \ref{s6}. We use the natural
unit of $c=\hbar=1$ through the paper.

\section {Model and Formalism}
\label{s2}
To have a general investigation, we consider the model Lagrangian
with two species of fermions interacting with each other via a
point interaction,
\begin{eqnarray}
{\cal
L}&=&\sum_{i=a,b}\psi_{i}^*(x)\left[-\frac{\partial}{\partial
\tau}+\frac{\nabla^2}{2m_i}+\mu_i\right]\psi_{i}(x)\nonumber\\
&+&g\psi_{a}^*( x)\psi_{b}^*(x)\psi_{b}( x)\psi_{a}(x)\
\end{eqnarray}
where $\psi_a(x),\psi_b(x)$ are fermion fields for the two species
$a$ and $b$ with space-time $x=(\tau,\vec{x})$, the coupling
constant $g$ is positive to keep the interaction attractive, $m_a$
and $m_b$ are the masses of the two species, and $\mu_a$ and
$\mu_b$ are the chemical potentials.

The key quantity to describe a thermodynamic system is the
partition function which can be defined as
\begin{equation}
Z=\int[d\psi_a][d\psi_a^*][d\psi_b][d\psi_b^*]e^{\int_0^\beta
d\tau\int d^3{\vec x}{\cal L}}
\end{equation}
in the imaginary time ($\tau$) formalism of finite temperature
field theory. According to the standard BCS theory, we introduce
the order parameter field $\Phi(x)$ and its complex conjugate
$\Phi^*(x)$,
\begin{equation}
\Phi(x)=g\big\langle\psi_{b}( x)\psi_{a}(x)\big\rangle\ ,\ \ \
\Phi^*(x)=g\big\langle\psi_{a}^*( x)\psi_{b}^*(x)\big\rangle\ ,
\end{equation}
where the symbol $\langle\ \rangle$ means ensemble average.
Introducing the Nambu-Gorkov space defined as
\begin{equation}
\Psi(x)=\left(\begin{array}{c} \psi_a(x)
\\ \psi_b^*(x)\end{array}\right)\ ,\ \ \ \Psi^*(x)=\left(\begin{array}{cc}
\psi_a^*(x)& \psi_b(x)\end{array}\right)\ ,
\end{equation}
the partition function in mean field approximation can be written
as
\begin{equation}
Z_{MF}=\int[d\Psi][d\Psi^*]e^{\int_0^\beta d\tau\int d^3{\vec
x}\left(\bar{\Psi}{\cal G}^{-1}\Psi+|\Phi|^2/g\right)}
\end{equation}
with the inverse of the mean field fermion propagator
\begin{equation}
{\cal G}^{-1}=\left(\begin{array}{cc} -\frac{\partial}{\partial
\tau}+\frac{\nabla^2}{2m_a}+\mu_a&\Phi(x)
\\ \Phi^*(x)&-\frac{\partial}{\partial
\tau}-\frac{\nabla^2}{2m_b}-\mu_b\end{array}\right)\ .
\end{equation}

To have a unified approach for the isotropic BCS, BP states and
anisotropic LOFF state, we take in the following a specific ansatz
for the order parameter fields
\begin{equation}
\Phi(x)=\Delta e^{2i\vec{q}\cdot\vec{x}}\ ,\ \ \ \Phi^*(x)=\Delta
e^{-2i\vec{q}\cdot\vec{x}}\ .
\end{equation}
Here $\Delta$ is a real quantity. Obviously, we recover the
isotropic state with $\vec{q}=0$ and obtain the anisotropic LOFF
state with $\vec{q}\neq 0$. After a transformation of the fermion
fields $\chi_a(x)=e^{i\vec{q}\cdot\vec{x}}\psi_a(x)$ and
$\chi_b(x)=e^{i\vec{q}\cdot\vec{x}}\psi_b(x)$, we can directly
evaluate the Gaussian path integral and obtain the thermodynamic
potential
\begin{equation}
\Omega(\Delta, \vec q)
=\frac{\Delta^2}{g}-T\sum_n\int\frac{d^3\vec{
p}}{(2\pi)^3}\textrm{Tr}\ln {\cal G}^{-1}(i\omega_n,\vec{p})
\end{equation}
in terms of the inverse fermion propagator in the momentum and
frequency space
\begin{equation}
{\cal G}^{-1}(i\omega_n,\vec{p})=\left(\begin{array}{cc}
i\omega_n-\epsilon_a&\Delta
\\ \Delta&i\omega_n+\epsilon_b\end{array}\right)
\end{equation}
with $\epsilon_a=\frac{(\vec p+\vec q)^2}{2m_a}-\mu_a$ and
$\epsilon_b=\frac{(\vec p-\vec q)^2}{2m_b}-\mu_b$. The explicit
form of the propagator can be written as
\begin{equation}
{\cal G}(i\omega_n,\vec{p})=\left(\begin{array}{cc} {\cal
G}_{11}(i\omega_n,\vec{p}) &{\cal G}_{12}(i\omega_n,\vec{p})
\\ {\cal G}_{21}(i\omega_n,\vec{p})&{\cal G}_{22}(i\omega_n,\vec{p})\end{array}\right)
\end{equation}
with the elements
\begin{eqnarray}
{\cal G}_{11}(i\omega_n,\vec{p}) &=&
{i\omega_n-\epsilon_-+\epsilon_+\over (i\omega_n-\epsilon_-)^2-\epsilon_\Delta^2},\nonumber\\
{\cal G}_{22}(i\omega_n,\vec{p}) &=&
{i\omega_n-\epsilon_--\epsilon_+\over
(i\omega_n-\epsilon_-)^2-\epsilon_\Delta^2},\nonumber\\
{\cal G}_{12}(i\omega_n,\vec{p}) &=& {-\Delta\over
(i\omega_n-\epsilon_-)^2-\epsilon_\Delta^2},\nonumber\\
{\cal G}_{21}(i\omega_n,\vec{p}) &=& {-\Delta\over
(i\omega_n-\epsilon_-)^2-\epsilon_\Delta^2},
\end{eqnarray}
where the quantities $\epsilon_+,\epsilon_-$ and $\epsilon_\Delta$
are defined as
\begin{eqnarray}
\epsilon_+&=&{\epsilon_a+\epsilon_b\over 2},\ \ \ \epsilon_-={\epsilon_a-\epsilon_b\over 2},\nonumber\\
\epsilon_\Delta&=&\sqrt{\epsilon_+^2+\Delta^2}.
\end{eqnarray}
Using the identity ${\bf Tr}\ln=\ln\det$, the thermodynamic
potential can be evaluated as
\begin{eqnarray}
\Omega(\Delta,\vec q)&=&\frac{\Delta^2}{g}-\int
\frac{d^3\vec{p}}{(2\pi)^3}\Big[\left(\frac{\epsilon_A}{2}+T\ln(1+e^{-{\epsilon_A\over T}})\right)\nonumber\\
&&+\left(\frac{\epsilon_B}{2}+T\ln(1+e^{-{\epsilon_B\over
T}})\right)-\epsilon_+\Big],
\end{eqnarray}
where $\epsilon_A$ and $\epsilon_B$ are the quasiparticle energies
\begin{eqnarray}
\epsilon_A=\epsilon_-+\epsilon_\Delta,\ \ \
\epsilon_B=\epsilon_--\epsilon_\Delta\ .
\end{eqnarray}

The gap equation which determines the order parameter $\Delta$
self-consistently is related to the off-diagonal element of the
propagator matrix,
\begin{equation}
\label{gap3} \Delta=gT\sum_n\frac{d^3\vec{p}}{(2\pi)^3}{\cal
G}_{12}(i\omega_n,\vec{p}),
\end{equation}
which is equivalent to the stationary condition of the
thermodynamic potential, $\partial \Omega/\partial \Delta=0$.
After the summation over the fermion frequency, it becomes
\begin{equation}
\label{gap4} \Delta(1-gI_\Delta)=0,
\end{equation}
where the function $I_\Delta$ is defined as
\begin{equation}
\label{idelta2}
I_\Delta=\int\frac{d^3\vec{p}}{(2\pi)^3}\frac{f(\epsilon_B)-f(\epsilon_A)}{2\epsilon_\Delta}
\end{equation}
with the Fermi-Dirac distribution function $f(x)=1/(e^{x/T}+1)$.
The physical pair momentum $\vec{q}$ is determined by minimizing
the thermodynamic potential. In the following, we choose a
suitable frame where the $z$ axis is along the direction of the
pair momentum, $\vec{q}=(0,0,q)$. Calculating the first order
derivative of the thermodynamic potential with respect to $q$, the
gap equation $\partial\Omega/\partial q=0$ for $q$ is explicitly
written as
\begin{eqnarray}
\label{gap5} 0&=&
\int\frac{d^3\vec{p}}{(2\pi)^3}\Bigg[\left(f(-\epsilon_B)-f(\epsilon_A)\right)\frac{\partial
\epsilon_-}{\partial
q}\nonumber\\
&&
-\left(1-\frac{\epsilon_+}{\epsilon_\Delta}+\left(f(\epsilon_A)+f(-\epsilon_B)\right)\frac{\epsilon_+}{\epsilon_\Delta}\right)\frac{\partial
\epsilon_+}{\partial q}\Bigg].
\end{eqnarray}
It is easy to see that $q=0$ is a trivial solution of the
equation, which corresponds to the isotropic BCS or BP state.

The occupation numbers are determined through the diagonal
elements of the fermion propagator,
\begin{eqnarray}
n_a(\vec{p})&=&T\lim_{\eta\to 0}\sum_n{\cal
G}_{11}(i\omega_n,\vec{p})e^{i\omega_n\eta},\nonumber\\
n_b(\vec{p})&=&-T\lim_{\eta\to 0}\sum_n{\cal
G}_{22}(i\omega_n,\vec{p})e^{-i\omega_n\eta}.
\end{eqnarray}
Completing the Matsubara frequency summation, we obtain
\begin{eqnarray}
n_a(\vec{p})&=&u_p^2f(\epsilon_A)
+v_p^2f(\epsilon_B),\nonumber\\
n_b(\vec{p})&=&u_p^2f(-\epsilon_B)+v_p^2f(-\epsilon_A)\
\end{eqnarray}
with the definition
$u_p^2=\left(1+\epsilon_+/\epsilon_\Delta\right)/2$ and
$v_p^2=\left(1-\epsilon_+/\epsilon_\Delta\right)/2$. At zero
temperature, they are reduced to
\begin{eqnarray}
\label{nab6}
n_a(\vec{p})&=&v_p^2\theta(\epsilon_\Delta-\epsilon_-)+u_p^2\theta(-\epsilon_\Delta-\epsilon_-),\nonumber\\
n_b(\vec{p})&=&v_p^2\theta(\epsilon_\Delta+\epsilon_-)+u_p^2\theta(-\epsilon_\Delta+\epsilon_-).
\end{eqnarray}
The particle number densities $n_a$ and $n_b$ can be calculated
from
\begin{equation}
n_a =\int{d^3\vec p\over (2\pi)^3}n_a(\vec p),\ \ \ n_b
=\int{d^3\vec p\over (2\pi)^3}n_b(\vec p).
\end{equation}
For a free Fermi gas, we can define the Fermi momenta $p_F^a$ and
$p_F^b$ for the two species, and the particle number densities are
then expressed as
\begin{equation}
n_a =\frac{(p_F^a)^3}{6\pi^2},\ \ \ n_b =\frac{(p_F^b)^3}{6\pi^2}.
\end{equation}
For convenience, we define total Fermi momentum $p_F$, Fermi
energy $\epsilon_F$ and Fermi velocity $v_F$ through the total
number density $n=n_a+n_b$ and the reduced mass
$m=2m_am_b/(m_a+m_b)$, $n=p_F^3/(3\pi^2), \epsilon_F=p_F^2/(2m)$
and $v_F=p_F/m$.

The above formalism is general for any system with $\mu_a\neq
\mu_b$ and $m_a\neq m_b$. In the following we will consider only
the case without mass difference between the two species,
$m_a=m_b=m$, which is relevant for the study of cold atomic
fermion gas, isospin asymmetric nuclear matter, and two flavor
color superconducting quark matter. For convenience, we define the
average of the chemical potentials $\mu_a$ and $\mu_b$ and their
difference,
\begin{equation}
\label{chemical} \mu=\frac{\mu_a+\mu_b}{2},\ \ \
\delta\mu=\frac{\mu_b-\mu_a}{2}.
\end{equation}
In the weak coupling BCS region, the chemical potentials $\mu_a$
and $\mu_b$ are approximately the corresponding Fermi energies
$\epsilon_F^a=(p_F^a)^2/(2m)$ and $\epsilon_F^b=(p_F^b)^2/(2m)$,
and the pairing gap $\Delta$ and the Fermi surface mismatch
$\delta\mu$ are much smaller than the average chemical potential,
$\Delta\sim\delta\mu\ll\mu$.

\subsection {Isotropic Superfluid with $\vec{q}=0$}
We discuss firstly the possible isotropic superfluid state, i.e.,
the case with $\vec{q}=0$. The dispersion relations of the
quasiparticle excitations are expressed as
\begin{eqnarray}
\label{dispersion2}
\epsilon_A&=&\delta\mu+\sqrt{\left(\frac{p^2}{2m}-\mu\right)^2+\Delta^2},\nonumber\\
\epsilon_B&=&
\delta\mu-\sqrt{\left(\frac{p^2}{2m}-\mu\right)^2+\Delta^2}.
\end{eqnarray}
Without losing generality, we set $\delta\mu\geq0$. It is easy to
see from the equation $\epsilon_B=0$ that only in the case with
$\Delta<\delta\mu$, there are gapless excitation at momenta
\begin{eqnarray}
\label{p12}
p_1&=&\sqrt{2m\left(\mu-\sqrt{\delta\mu^2-\Delta^2}\right)},\nonumber\\
p_2&=&\sqrt{2m\left(\mu+\sqrt{\delta\mu^2-\Delta^2}\right)}.
\end{eqnarray}
In the gapless state, we have $n_a(\vec{p})= n_b(\vec{p})=v_p^2$
for $p<p_1$ and $p>p_2$ and $n_a(\vec{p})=0, n_b(\vec{p})=1$ for
$p_1<p<p_2$, the superfluidity state with fermion pairing in the
regions $p<p_1$ and $p>p_2$ is breached by the normal fermion gas
in the region $p_1<p<p_2$. This is the reason why people call the
gapless state as breached pairing (BP) state. In the case with
$\Delta>\delta\mu$, there is no gapless excitation, and the system
is in BCS phase with $n_a(\vec{p})= n_b(\vec{p})=v_p^2$ for all
$p$.

\subsection {Anisotropic Superfluid with $\vec{q}\neq 0$}
We now discuss the anisotropic superfluid state with
$\vec{q}\neq0$. The dispersion relations of the elementary
excitations can be written as
\begin{eqnarray}
\label{energy4}
\epsilon_A&=&\delta\mu+\frac{pq}{m}\cos\theta+\sqrt{\left(\frac{p^2+q^2}{2m}-\mu\right)^2+\Delta^2},\nonumber\\
\epsilon_B&=&
\delta\mu+\frac{pq}{m}\cos\theta-\sqrt{\left(\frac{p^2+q^2}{2m}-\mu\right)^2+\Delta^2},
\end{eqnarray}
where $\theta$ is the angle between the momenta $\vec p$ and $\vec
q$. In many of the previous investigations, the term $q^2/2m$ was
neglected and the term $pq\cos\theta/m$ was approximated by
$v_Fq\cos\theta$\cite{larkin,fulde,takada} since $q/p_F\ll 1$ in
weak coupling limit. Taking this approximation, in the case of
$\delta_\theta=\delta\mu+v_Fq\cos\theta>\Delta$, the gapless
excitation happens at
\begin{eqnarray}
\label{p12}
p_1(\theta)&=&\sqrt{2m\left(\mu-\sqrt{\delta_\theta^2-\Delta^2}\right)},\nonumber\\
p_2(\theta)&=&\sqrt{2m\left(\mu+\sqrt{\delta_\theta^2-\Delta^2}\right)}.
\end{eqnarray}
For the following numerical research, we will do full calculation
without employing this approximation. We will see that the
anisotropic LOFF state possesses always unequal particle numbers
for the two species, and hence the LOFF superfluid is gapless.

Since the gap equation for $\Delta$ suffers ultraviolet
divergence, a regularization scheme is needed. We consider now two
regularization schemes. In the first scheme, we simply employ a
hard cutoff $p_\Lambda$ for the three momentum. Such a
regularization scheme is related to the study of color
superconductivity at moderate baryon density and isospin
asymmetric nuclear matter. To avoid a specific choice of model
parameters, such as the mass $m$, we express the coupling $g$ in
terms of the s-wave scattering length $a_s$ as $g=\frac{4\pi
a}{m}$ with $a=|a_s|$. Once this is done, the two original
parameters $p_\Lambda$ and $g$ in the model are replaced by two
dimensionless parameters $\Lambda=(p_\Lambda/p_F)^2$ and $p_Fa$.
In the weak coupling BCS region $0<p_Fa<1$ and for not very large
cutoff $1<\Lambda<2$, the solution of the gap equation can be well
described by
\begin{equation}
\Delta_0\simeq {8\over e^2}\epsilon_F\sqrt{\Lambda^2-1}\
e^{-\frac{\pi}{2p_Fa}}
\end{equation}
in the symmetric case with $\delta\mu=0$.

In the second regularization scheme, we replace the bare coupling
$g$ by the low energy limit of the two-body $T$-matrix\cite{pao}
\begin{equation}
\frac{m}{4\pi a_s}=-\frac{1}{g}+\int\frac{d^3{\vec
p}}{(2\pi)^3}\frac{m}{p^2}.
\end{equation}
Such a regularization scheme is often used in the study of atomic
fermion gas. In this scheme, it is not necessary to introduce a
momentum cutoff, and all physical results depend only on the
dimensionless coupling $p_Fa$. This can be seen explicitly by
scaling all the quantities with energy dimension by the Fermi
energy $\epsilon_F$ and all the quantities with momentum dimension
by the Fermi momentum $p_F$,
\begin{eqnarray}
\label{dless}
&& z=\left({p\over p_F}\right)^2,\ z_q=\left({q\over
p_F}\right)^2,\ \nu_a={\mu_a\over \epsilon_F},\ \nu_b={\mu_b\over \epsilon_F},\nonumber\\
&& \nu={\mu\over \epsilon_F},\ \ \delta\nu={\delta\mu\over
\epsilon_F},\ \ \delta={\Delta\over \epsilon_F},\ \ t={T\over
\epsilon_F}.
\end{eqnarray}
With these dimensionless quantities, we obtain the scaled gap
equations,
\begin{eqnarray}
\label{gap9}
0 &=&\delta\left[\int dz
\sqrt{z}\int_{-1}^1dx\frac{f(\epsilon_{z,x}^B)-f(\epsilon_{z,x}^A)}{\sqrt{(z+z_q-\nu)^2+\delta^2}}-\frac{2\pi}{p_Fa}\right],\nonumber\\
0 &=&\int dz
\sqrt{z}\int_{-1}^1dx\Bigg[x\sqrt{\frac{z}{z_q}}\left(f(\epsilon_{z,x}^A)+f(\epsilon_{z,x}^B)\right)\nonumber\\
& +
&\frac{z+z_q-\nu}{\sqrt{(z+z_q-\nu)^2+\delta^2}}\left(f(\epsilon_{z,x}^A)-f(\epsilon_{z,x}^B)\right)+1\Bigg],
\end{eqnarray}
the scaled thermodynamic potential,
\begin{eqnarray}
\label{somega2} \frac{\Omega}{E_0} &=& -{1\over 2}\int
dz\sqrt{z}\int_{-1}^1dx\Bigg[\left(\frac{\epsilon_{z,x}^A}{2}+t\ln(1+e^{-{\epsilon_{z,x}^A\over
t}})\right)\nonumber\\
&+&\left(\frac{\epsilon_{z,x}^B}{2}
+t\ln(1+e^{-{\epsilon_{z,x}^B\over
t}})\right)-(z+z_q-\nu)\Bigg]\nonumber\\
&+&\frac{\pi \delta^2}{2p_Fa},
\end{eqnarray}
and the scaled number densities,
\begin{eqnarray}
\label{sn2}
\frac{n_a}{n_0} &=& {1\over 2}\int dz
\sqrt{z}\int_{-1}^1 dx
\left[u_z^2f(\epsilon_{z,x}^A)+v_z^2f(\epsilon_{z,x}^B)\right],\\
\frac{n_b}{n_0} &=& {1\over 2}\int dz \sqrt{z}\int_{-1}^1 dx
\left[1-u_z^2f(\epsilon_{z,x}^B)-v_z^2f(\epsilon_{z,x}^A)\right],\nonumber
\end{eqnarray}
where the normalization constants $n_0$ and $E_0$ are chosen to be
$n_0=p_F^3/(4\pi^2)$ and $E_0=p_F^5/(8\pi^2m)$, the quasiparticle
energies and the functions $u, v$ are reexpressed in terms of the
dimensionless quantities,
\begin{eqnarray}
\label{red}
&&
\epsilon_{z,x}^A=\delta\nu+2x\sqrt{zz_q}+\sqrt{\left(z+z_q-\nu\right)^2+\delta^2},\\
&& \epsilon_{z,x}^B=\delta\nu+2x\sqrt{zz_q}-\sqrt{\left(z+z_q-\nu\right)^2+\delta^2},\nonumber\\
&&
u_z^2=\frac{1}{2}\left(1+(z+z_q-\nu)/\sqrt{(z+z_q-\nu)^2+\delta^2}\right),\nonumber\\
&&
v_z^2=\frac{1}{2}\left(1-(z+z_q-\nu)/\sqrt{(z+z_q-\nu)^2+\delta^2}\right).\nonumber
\end{eqnarray}
In weak coupling limit, the solution of the gap equation is
\begin{equation}
\Delta_0\simeq {8\over e^2}\epsilon_F e^{-\frac{\pi}{2p_Fa}}
\end{equation}
in the symmetric case with $\delta\mu=0$. It is
obvious that if we take $\Lambda=\sqrt{2}$, the results in the two
regularization schemes are the same in symmetric case and will not
make significant difference in asymmetric case in the region
$0<p_Fa<1$.

While the values of $\Delta,\delta\mu$ and $q$ depend on the
cutoff $\Lambda$, we emphasis that the scaled quantities such as
$\Delta/\Delta_0, \delta\mu/\Delta_0$ and $v_Fq/\delta\mu$ are
regularization scheme independent in the weak coupling region
\cite{bedaque,caldas,giannakis2,gubankova}.

\section {Thermodynamic and Dynamical Instabilities in BP Superfluid}
\label{s3}
In this section, we discuss some possible thermodynamic and
dynamical instabilities of the BP state. From the analytic
analysis, we will see that there are two types of instabilities in
the BP phase. One is the Sarma instability which was found many
years ago, and the other is the magnetic instability which was
recently found in the study of gapless superconductivity.

We first discuss the thermodynamic instability induced by
condensate fluctuations. To this end, we expand the thermodynamic
potential in powers of an infinitely small fluctuation $\delta$,
\begin{equation}
\label{delta1}
\Omega(\Delta+\delta)-\Omega(\Delta)=\frac{\partial
\Omega}{\partial \Delta}\delta+ \frac{1}{2}\frac{\partial^2
\Omega}{\partial \Delta^2}\delta^2+\cdots.
\end{equation}
The linear term on the r.h.s. vanishes automatically due to the
gap equation for $\Delta$. At zero temperature, the gap
susceptibility can be evaluated as
\begin{eqnarray}
\label{kd3}
\kappa_\Delta &=& {\partial^2 \Omega\over \partial
\Delta^2}\\
&=&\int_0^\infty\frac{dpp^2}{2\pi^2}\frac{\Delta^2}{\epsilon^2_\Delta}\left[\frac{\theta(\epsilon_\Delta-\delta\mu)}{\epsilon_\Delta}
-\delta(\epsilon_\Delta-\delta\mu)\right].\nonumber
\end{eqnarray}
In weak coupling region with $\Delta\sim\delta\mu\ll\mu$, it can
be approximated as
\begin{equation}
\kappa_\Delta\approx\frac{mp_F}{\pi^2}\left[1-
\frac{\delta\mu\theta(\delta\mu-\Delta)}{\sqrt{\delta\mu^2-\Delta^2}}\right].
\end{equation}

For the BCS state which satisfies the constraint
$\Delta>\delta\mu$, the $\delta$-function in (\ref{kd3})
disappears automatically, $\kappa_\Delta$ is obviously positive,
and the state is stable. For the BP state with $\Delta <
\delta\mu$, the integral over the $\delta$-function becomes
nonzero and dominates the susceptibility. In this case, we have
$\kappa_\Delta < 0$, and the state is unstable. This is the
so-called Sarma instability.

For the systems with fixed chemical potentials $\mu_a$ and $\mu_b$
or fixed total chemical potential $\mu$ and chemical potential
difference $\delta\mu$, we can directly compare $\delta\mu$ with
$\Delta$ and see if the state is stable. However, for a systems
under some constraints on particle numbers, the grand canonical
ensemble should be replaced by canonical ensemble, and the
essential quantity to describe the thermodynamics of the systems
is the free energy $F$ instead of the thermodynamic potential
$\Omega$. In this case, the chemical potentials are no longer
independent thermodynamic variables, $\Delta$ and $\delta\mu$
should be determined self-consistently via solving the coupled set
of equations for the gap parameter and the constraints on the
particle numbers. Therefore, it is not convenient to judge the
instability of the BP state through $\kappa_\Delta$. On the other
hand, when the number difference between the two species is fixed,
the BCS solution is ruled out, and only the BP state
survives\cite{shovkovy,huang2}. A direct way for the instability
analysis is to compare the thermodynamic potentials or free
energies themselves in BP and LOFF states, and see which one is
more stable.

We now turn to discuss the instability induced by a small
superfluid velocity $\vec{v}_s$. When the superfluid moves with a
uniform velocity $\vec{v}_s$, the condensate transforms like
$\Phi\rightarrow\Phi e^{2im\vec{v}_s\cdot\vec{x}}$ and
$\Phi^*\rightarrow\Phi^* e^{-2im\vec{v}_s\cdot\vec{x}}$, and the
supercurrent $\vec{j}_s$ and the superfluid density $\rho_s$ are
defined via the thermodynamic potential
\begin{equation}
\Omega(\vec{v}_s)=\Omega(\vec{v}_s=0)+\vec{j}_s\cdot\vec{v}_s+\frac{1}{2}\rho_s\vec{v}_s^2+\cdots.
\end{equation}
When the density $\rho_s$ is negative, the superfluid is
dynamically unstable. This dynamical instability can be related to
the thermodynamic instability induced by pair momentum
fluctuations. We perturb the BP state with an infinitely small
LOFF momentum $q$ and expand the thermodynamic potential around
$q=0$,
\begin{equation}
\label{kq2} \Omega(\Delta,q)-\Omega(\Delta,0)=\frac{\partial
\Omega}{\partial q}\bigg|_{q=0}q+\frac{1}{2}\frac{\partial^2
\Omega}{\partial q^2}\bigg|_{q=0}q^2+\cdots.
\end{equation}
The linear term vanishes due to the gap equation for the pair
momentum, and we can easily observe the relations between the
superfluid density $\rho_s$\cite{wu}, the Meissner mass squared
$m_A^2$\cite{he} and the pair momentum
susceptibility\cite{giannakis} $\kappa_q =
\partial^2 \Omega/\partial q^2|_{q=0}$,
\begin{equation}
\label{kq1} \rho_s=m^2\kappa_q,\ \ m_A^2=e^2\kappa_q.
\end{equation}
Using the result in \cite{he}, at zero temperature we have
\begin{equation}
\label{kq3}
\rho_s=mn-\frac{1}{6\pi^2}\int_0^\infty
dpp^4\delta(\epsilon_\Delta-\delta\mu).
\end{equation}
In the weak coupling region, $\epsilon_\Delta-\delta\mu=0$ has two
possible roots $p_1$ and $p_2$, and $\rho_s$ is reduced to
\begin{equation}
\label{kq4}
\rho_s=mn\left[1-\eta\frac{\delta\mu\theta(\delta\mu-\Delta)}{\sqrt{\delta\mu^2-\Delta^2}}\right]
\end{equation}
with the coefficient
$\eta=\left(p_1^3+p_2^3\right)/\left(6\pi^2n\right)$. Since the
total fermion density $n$ at $T=0$ can be expressed as
\begin{equation}
\label{density}
n=\frac{p_1^3+p_2^3}{6\pi^2}+\frac{1}{\pi^2}\left(-\int_0^{p_1}dpp^2u_p^2+\int_{p_2}^\infty
dpp^2v_p^2\right),
\end{equation}
and the two integrals in the bracket are very small in weak
coupling region and thus can be neglected\cite{he}, the
coefficient $\eta$ is approximately equal to $1$, and $\rho_s$ can
be expressed as
\begin{equation}
\label{kq5}
\rho_s \approx
mn\left[1-\frac{\delta\mu\theta(\delta\mu-\Delta)}{\sqrt{\delta\mu^2-\Delta^2}}\right].
\end{equation}
It is easy to see that in the BP phase with $\Delta<\delta\mu$,
$\rho_s$ becomes negative. This is the so called magnetic
instability, since it is directly related to the negative Meissner
mass squared, if the fermions are charged. This dynamical
instability implies a lower free energy of the LOFF state in
comparison with the BP state.

As we mentioned above, however, for the systems with fixed density
asymmetry instead of chemical potential asymmetry, the essential
thermodynamic function is the free energy $F$ instead of $\Omega$.
For such systems, the above argument fails and we need to
calculate the free energy directly and then determine the
thermodynamically stable state.

We end this section with a qualitative discussion on the
instability analysis in strong coupling case. For such systems,
the condition $\mu>>\delta\mu$ does not hold and even $\mu<0$
happens in the BEC region, the node $p_1$ becomes imaginary, and
there is only one gapless node $p_2$. In this case, it has been
argued that the gapless state is automatically free from the Sarma
and magnetic instabilities\cite{pao,kitazawa}, and therefore, the
LOFF state becomes less favored than the BP state. In fact, some
recent studies have shown that the BP state will be the ground
state of asymmetric fermion gas when the coupling is
strong\cite{son,sheehy,yang2,carlson}.

\section {Fixed Chemical Potential Asymmetry}
\label{s4}
We now solve the gap equations to find the thermodynamically
stable state among the normal, BCS, BP and LOFF states. The four
states are defined through the solutions of the gap equations,\\
(1)normal phase with $\Delta=0$,\\
(2)BCS phase with $\Delta\neq0, q=0$ and $\Delta>\delta\mu$,\\
(3)BP phase with $\Delta\neq0, q=0$ and $\Delta<\delta\mu$,\\
(4)LOFF phase with $\Delta\neq0, q\neq0$. \\

\begin{figure}
\centering
\includegraphics[width=6cm]{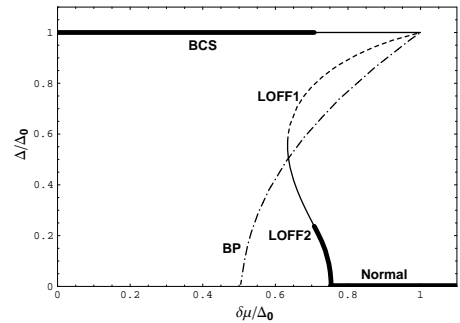}
\caption{The scaled gap parameter $\Delta/\Delta_0$ as a function
of the scaled chemical potential difference $\delta\mu/\Delta_0$
at $p_F a =0.4$ and $T=0$ for normal, BCS, BP and LOFF (including
LOFF1 by dashed line and LOFF2 by solid line) states. The thick
lines indicate the real ground states in different regions.}
\label{fig1}
\end{figure}

\begin{figure}[!htb]
\begin{center}
\includegraphics[width=6cm]{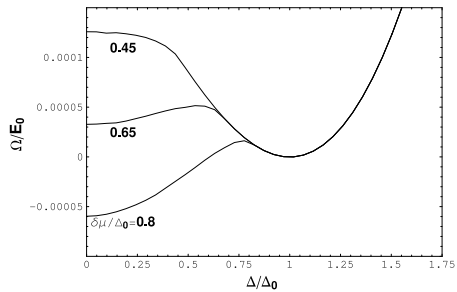}
\caption{The scaled thermodynamic potential $\Omega/E_0$ as a
function of the scaled gap parameter $\Delta/\Delta_0$ at $p_F a
=0.4, T=0$ and $q=0$ for three chemical potential values,
$0<\delta\mu/\Delta_0<1/2, 1/2<\delta\mu/\Delta_0<1/\sqrt 2$ and
$\delta\mu/\Delta_0>1/\sqrt 2$. We have chosen the thermodynamic
potential of the BCS state to be zero.} \label{fig2}
\end{center}
\end{figure}

In this section we consider the case of fixed chemical potentials
$\mu_a$ and $\mu_b$ with $\mu_a\neq\mu_b$. For example, if a
superfluid with cooper pair composed of spin up and down fermions
is placed in an external magnetic field $H$, there will be an
effective chemical potential asymmetry $\delta\mu=\mu_BH$ with the
fermion magneton $\mu_B$. When the external field is fixed, the
chemical potential asymmetry is naturally fixed. In this case, the
system is connected to reservoirs of each species and the particle
numbers of the two species are not conserved. Fig.\ref{fig1} shows
the condensate $\Delta$, scaled by its value $\Delta_0$ in the
corresponding symmetric system with $\delta\mu=0$, as a function
of the chemical potential difference $\delta\mu$ scaled by
$\Delta_0$ too at $p_Fa=0.4$ and $T=0$. In a wide range of
coupling $0<p_Fa<1$, we found that the results are almost the same
for the scaled quantities. We compare first the three uniform
phases. The solution $\Delta=0$ is the trivial solution of the gap
equations which describes the normal phase without superfluidity,
the BCS solution does not change with increasing $\delta\mu$ and
keeps its symmetric value until $\delta\mu=\Delta_0$, and the BP
solution starts to appear at $\delta\mu=0.5\Delta_0$ and satisfies
$\Delta<\delta\mu$ in the region $0.5\Delta_0<\delta\mu<\Delta_0$.
The BP solution can be expressed analytically as
$\Delta=\sqrt{\Delta_0(2\delta\mu-\Delta_0)}$\cite{sarma}. From
the analysis in Section \ref{s3}, the BP state is
thermodynamically unstable. This can be seen clearly from the
thermodynamic potential as a function of the scaled condensate
$\Delta/\Delta_0$ for three values of scaled chemical potential
difference $\delta\mu/\Delta_0$, shown in Fig.\ref{fig2}. For
$\delta\mu<0.5\Delta_0$, there are a maximum at $\Delta=0$ and a
minimum at $\Delta=\Delta_0$, corresponding, respectively, to the
normal phase and BCS phase, and for
$0.5\Delta_0<\delta\mu<0.707\Delta_0\simeq\Delta_0/\sqrt 2$, there
are a maximum at $0<\Delta<\Delta_0$, a local minimum at
$\Delta=0$ and the lowest minimum at $\Delta=\Delta_0$,
corresponding, respectively, to the BP state, normal state and BCS
state. For $\delta\mu>\Delta_0/\sqrt 2$, the maximum is still
located at $0<\Delta<\Delta_0$, but the local minimum and the
lowest minimum exchange positions. Therefore, among the three
uniform phases the BCS phase is stable in the region
$0<\delta\mu<\Delta_0/\sqrt 2$ and the normal phase becomes stable
for $\delta\mu>\Delta_0/\sqrt 2$.
\begin{figure}[!htb]
\begin{center}
\includegraphics[width=6cm]{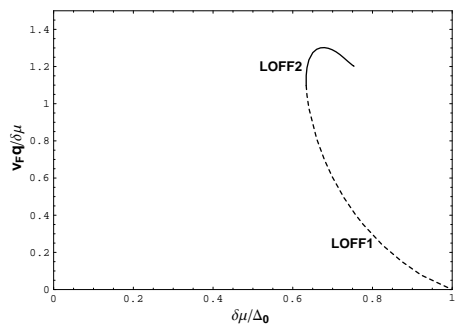}
\caption{The scaled LOFF momentum $v_Fq/\delta\mu$ as a function
of the scaled chemical potential difference $\delta\mu/\Delta_0$
at $p_F a =0.4$ and $T=0$ for LOFF1 and LOFF2.} \label{fig3}
\end{center}
\end{figure}
\begin{figure}[!htb]
\begin{center}
\includegraphics[width=6cm]{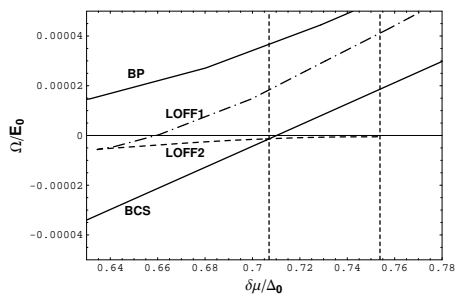}
\caption{The scaled thermodynamic potential $\Omega/E_0$ as a
function of the scaled chemical potential difference
$\delta\mu/\Delta_0$ at $p_F a =0.4$ and $T=0$ for normal, BCS,
BP, LOFF1, and LOFF2 states. We have chosen the thermodynamic
potential of the normal state to be zero.} \label{fig4}
\end{center}
\end{figure}
\begin{figure}[!htb]
\begin{center}
\includegraphics[width=6cm]{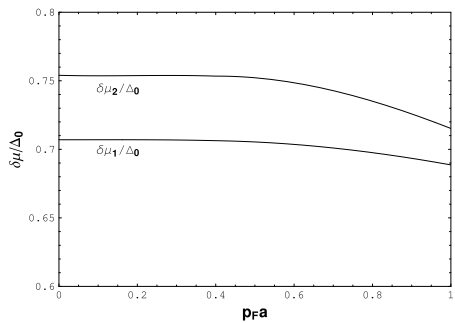}
\caption{The starting and ending chemical potential difference
$\delta\mu_1/\Delta_0$ and $\delta\mu_2/\Delta_0$ for the LOFF
window as functions of $p_F a$. } \label{fig5}
\end{center}
\end{figure}

We now include the anisotropic LOFF state. The scaled LOFF gap
$\Delta/\Delta_0$ and scaled LOFF momentum $v_Fq/\delta\mu$ are
shown in Figs.\ref{fig1} and \ref{fig3} as functions of the scaled
chemical potential difference $\delta\mu/\Delta_0$ at $p_F a =0.4$
and $T=0$. The LOFF state starts at $\delta\mu=0.635\Delta_0$ and
there are two branches called LOFF1 and LOFF2. To find the true
ground state, we compared the thermodynamic potentials for the
four phases in Fig.\ref{fig4}. It is easy to see that, below
$\delta\mu_1\simeq\Delta_0/\sqrt 2=0.707\Delta_0$ the LOFF1 and
LOFF2 are unstable and the system is in BCS state, above
$\delta\mu_2\simeq 0.754\Delta_0$ the normal state becomes stable,
and in between $\delta\mu_1$ and $\delta\mu_2$ the LOFF2 state is
the true ground state. In this LOFF window, the gap $\Delta$
starts from $0.24\Delta_0$ at $\delta\mu=\delta\mu_1$ and
decreases monotonically to $0$ at $\delta\mu=\delta\mu_2$, the
scaled LOFF monentum $v_Fq/\delta\mu$ is almost a constant,
$1.2<v_Fq/\delta\mu<1.3$ in the window. The critical values
$\delta\mu_1$ and $\delta\mu_2$ for the LOFF state agree with the
analytical and numerical results obtained in other studies in weak
coupling
limit\cite{larkin,fulde,takada,alford5,casalbuoni4,giannakis2}. In
Fig.\ref{fig5} we showed $\delta\mu_1/\Delta_0$ and
$\delta\mu_2/\Delta_0$ as functions of $p_F a$. In small coupling
region, they are almost constants and agree well with the
analytical result, while when the coupling is strong enough, the
LOFF window becomes small. This behavior agrees with the result in
crystalline color superconductivity\cite{alford5}.

We investigated also systems under condition of fixed total number
$n$ and chemical potential asymmetry $\delta\mu$. The behavior of
the gap parameter, pair momentum and the free energy $F=\Omega+\mu
n$ is almost the same as what we showed here. The reason may be
the assumption $m_a=m_b$ considered in this paper. With a large
mass difference, the two cases will be quite
different\cite{gubankova}.
\section {Fixed Density Asymmetry }
\label{s5}
For many asymmetric fermion systems, the fixed quantity is the
number density asymmetry instead of the chemical potential
asymmetry. For example, in cold atom gas of $^6Li$ the number of
the atoms in each hyperfine state can be directly adjusted, in
isospin asymmetric nuclear matter the density difference between
neutron and proton is fixed, and in two flavor dense quark matter,
the density difference between u and d quarks is required to
satisfy charge neutrality. In this section, we assume that the
number densities of the two species are fixed. In this case, the
chemical potentials $\mu_a$ and $\mu_b$ are not thermodynamic
variables, they should be determined by the known number densities
$n_a$ and $n_b$. As a result, the essential quantity to describe
the thermodynamics of the system is the free energy defined as
\begin{equation}
\label{free2}
F=\Omega+\mu_a n_a + \mu_b n_b.
\end{equation}

In general case of $n_a\ne n_b$, instead of $n_a$ and $n_b$, we
can equivalently describe the system by the total number
$n=n_a+n_b$ and the quantity
\begin{equation}
\label{alpha}
\alpha={n_b-n_a\over n_b+n_a}
\end{equation}
which controls the degree of asymmetry between the two species and
satisfies $0<\alpha<1$. Obviously, once $n_a\ne n_b$ or $\alpha\ne
0$, the BCS pairing mechanism is ruled out, only the normal phase,
BP phase and LOFF phase are left as candidates of ground states.

Considering the coupled set of equations for the gap parameter and
the particle numbers
\begin{equation}
\label{nanb} n_a={1-\alpha\over 2}n,\ \ \ n_b={1+\alpha\over 2}n,
\end{equation}
the gap parameter $\Delta$ and chemical potential difference
$\delta\mu$ in the BP state as functions of the asymmetry degree
$\alpha$ are shown in Fig.{\ref{fig6} at $T=0$, where the scaling
parameter $\alpha_c$ is the critical asymmetry where the BP
superfluidity disappears. While both $\Delta$ and $\delta\mu$ drop
down with increasing degree of asymmetry, they always satisfy the
condition for the BP state, $\Delta<\delta\mu$, and the difference
between them becomes larger and larger. Finally, they end at the
critical value $\alpha_c$. From the comparison with Fig.\ref{fig1}
where $\delta\mu$ is fixed, the BP gap parameter here still starts
with $\Delta=0$ at $\delta\mu=0.5\Delta_0$ and ends with
$\Delta=\Delta_0$ at $\delta\mu=\Delta_0$. The $\alpha$-dependence
of $\Delta$ and $\delta\mu$ and the critical value $\alpha_c$ in
our calculation agree well with the analytic results obtained in
weak coupling region\cite{petit,lombardo},
\begin{eqnarray}
\label{limit}
\frac{\Delta(\alpha)}{\Delta_0}=\sqrt{1-\frac{\alpha}{\alpha_c}},\
\ \
\frac{\delta\mu(\alpha)}{\Delta_0}=1-\frac{1}{2}\frac{\alpha}{\alpha_c}.
\end{eqnarray}

The question is which of the normal and BP states is relatively
stable. From the direct calculation of the free energy as a
function of the gap parameter for a specific asymmetry
$\alpha=0.05$ and coupling $p_F a =0.8$ at $T=0$, shown in
Fig.\ref{fig7}, the trivial solution $\Delta=0$ is the maximum but
the BP solution $\Delta\ne 0$ is the minimum of the free energy.
We have checked this result for other values of $\alpha$ and
$p_Fa$, the qualitative behavior does not change. Therefore, the
Sarma instability is avoided and the BP state becomes a stable
solution in the family of isotropic phases. It is necessary to
emphasis that the point to avoid the Sarma instability is the
constraint of unequal numbers of the two species.
\begin{figure}
\centering
\includegraphics[width=6cm]{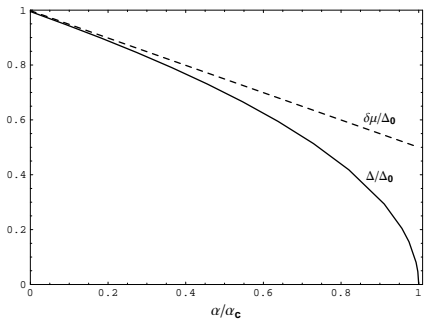}
\caption{The scaled gap parameter $\Delta/\Delta_0$ and the
chemical potential difference $\delta\mu/\Delta_0$ as functions of
the scaled degree of asymmetry $\alpha/\alpha_c$ at $T=0$.
$\alpha_c$ is the critical asymmetry for the BP superfluidity.}
\label{fig6}
\end{figure}
\begin{figure}
\centering
\includegraphics[width=6cm]{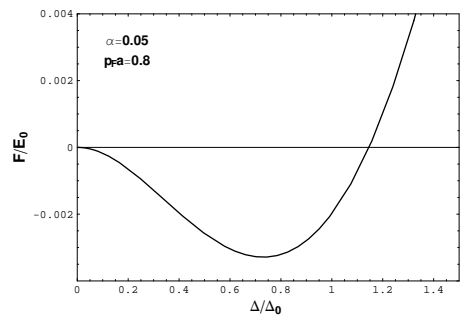}
\caption{The scaled free energy $F/E_0$ as a function of the
scaled gap parameter $\Delta/\Delta_0$ at $T=0, \alpha=0.05$ and
$p_F a =0.8$. The minimum and maximum correspond, respectively, to
the BP and normal states. } \label{fig7}
\end{figure}
\begin{figure}
\centering
\includegraphics[width=6cm]{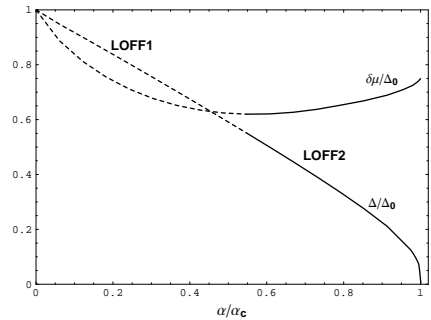}
\caption{The scaled LOFF gap $\Delta/\Delta_0$ and chemical
potential difference $\delta\mu/\Delta_0$ as functions of the
scaled degree of asymmetry $\alpha/\alpha_c$ at $T=0$. The dashed
and solid lines correspond, respectively, to the branches LOFF1
and LOFF2.} \label{fig8}
\end{figure}

The LOFF solution is obtained through solving the equations for
the gap $\Delta$ and LOFF momentum, together with the known
particle number densities $n_a$ and $n_b$. At zero temperature,
the scaled gap parameter and LOFF momentum are plotted as
functions of the scaled degree of asymmetry $\alpha/\alpha_c$ in
Figs.\ref{fig8} and \ref{fig9}. Here $\alpha_c$ is the critical
asymmetry where the LOFF superfluidity disappears. We have checked
that for any $\alpha<\alpha_c$, the LOFF solution obtained here is
thermodynamically stable against a small perturbation of $\Delta$,
i.e., it corresponds to the minimum of $F$ as a function of
$\Delta$. Exactly the same as that shown in Fig.\ref{fig1}, the
two branches LOFF1 and LOFF2 meet at $\delta\mu=0.635\Delta_0$,
LOFF1 ends at $\Delta=\Delta_0$ and $\delta\mu=\Delta_0$, and
LOFF2 ends at $\Delta=0$ and $\delta\mu\simeq 0.754\Delta_0$. If
we denote the asymmetry corresponding to the meeting point of
LOFF1 and LOFF2 as $\alpha_0$, we found
$\alpha_0\simeq0.55\alpha_c$. The scaled LOFF momentum can vary in
the range $0<v_Fq/\delta\mu<1.3$. In Fig.\ref{fig10}, we showed
the critical asymmetry for BP and LOFF states as a function of
coupling strength $p_Fa$. We found that the critical asymmetry of
LOFF state is larger than that of BP state, and the difference
between the two increases with increasing coupling strength. It is
necessary to note that the condition for the BP gap parameter,
$\Delta<\delta\mu$, is no longer necessary for the LOFF gap
parameter, because of the term $pq\cos\theta/m$ in the dispersion
relation (\ref{energy4}).

\begin{figure}
\includegraphics[width=6cm]{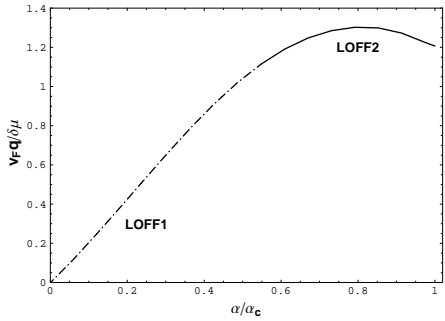}
\caption{The scaled LOFF momentum $v_Fq/\delta\mu$ as a function
of the scaled degree of asymmetry $\alpha/\alpha_c$ at $T=0$. The
dashed and solid lines correspond, respectively, to the branches
LOFF1 and LOFF2.} \label{fig9}
\end{figure}

\begin{figure}[!htb]
\begin{center}
\includegraphics[width=6cm]{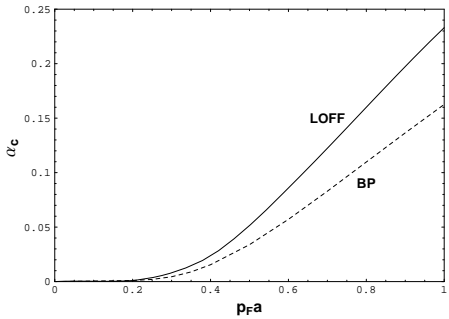}
\caption{The critical asymmetry $\alpha_c$ as a function of the
interaction strength $p_F a$. The solid and dashed lines
correspond, respectively, to the LOFF and BP states. }
\label{fig10}
\end{center}
\end{figure}
\begin{figure}
\centering
\includegraphics[width=6cm]{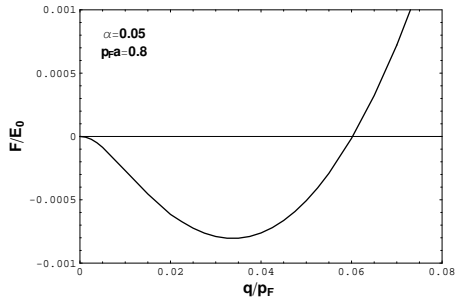}
\caption{The scaled free energy $F/E_0$ as a function of the
scaled pair momentum $q/p_F$ at $T=0, \alpha=0.05$ and $p_Fa=0.8$.
The maximum at zero pair momentum and the minimum at finite pair
momentum correspond, respectively, to the BP and LOFF states. }
\label{fig11}
\end{figure}
\begin{figure}
\centering
\includegraphics[width=6cm]{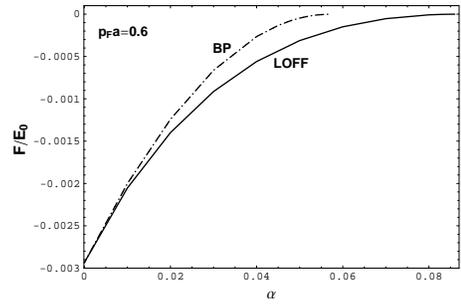}
\includegraphics[width=6cm]{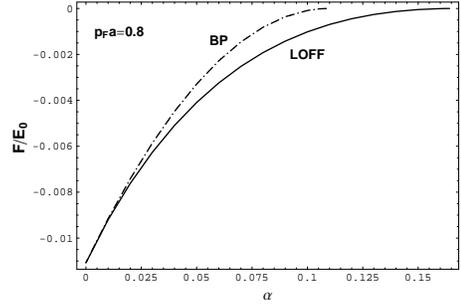}
\caption{The scaled free energies for the normal, BP and LOFF
states as functions of the degree of asymmetry $\alpha$ at $T=0$
and $p_F a=0.6$ and $0.8$. We have chosen the free energy of the
normal state to be zero.} \label{fig12}
\end{figure}
\begin{figure}
\centering
\includegraphics[width=6cm]{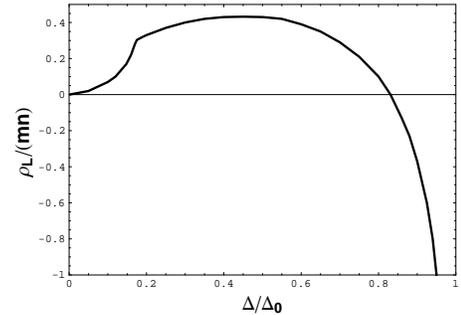}
\caption{The logitudinal superfluid density, scaled by $mn$, as a
function of $\Delta/\Delta_0$. } \label{fig13}
\end{figure}

Now the question left is to extract the real ground state from the
isotropic and anisotropic phases. Since the BCS phase is already
kicked out, we need to compare the free energies only for the
normal, BP and LOFF states. Fig.\ref{fig11} shows the free energy
as a function of the pairing momentum at fixed degree of asymmetry
and coupling strength. It is clear that the BP state which has a
zero pair momentum is a maximum of the free energy in the $q$
direction, while the LOFF state which has a finite pair momentum
is the minimum of the free energy. We note that, the LOFF minimum
in the figure locates at the LOFF1 branch which is ruled out in
the conventional case studied in Section \ref{s4}. For other
values of asymmetry and coupling, the $F-q$ curve behaves
similarly. In Fig.\ref{fig12}, we calculated the free energies for
normal, BP and LOFF states as functions of $\alpha$ at two values
of the coupling strength, $p_Fa=0.6$ and $0.8$. For other values
of coupling in the region $0<p_Fa<1$ we have checked that the
qualitative behavior is the same. Thus we have the conclusion
that, in the region with small asymmetry where the BP and LOFF
states coexist, the LOFF is the stable one, and in the region with
higher asymmetry where the BP disappears, the LOFF becomes the
only possible mechanism for superfluidity. It is necessary to
point out that, while only a part of LOFF2 is thermodynamically
stable in the case with fixed chemical potential asymmetry
$\delta\mu$, both LOFF1 and LOFF2 are thermodynamically stable in
the case with fixed density asymmetry $\alpha$. The conventional
LOFF window $0.707\Delta_0<\delta\mu<0.754\Delta_0$ corresponds to
the highly asymmetric region $0.93\alpha_c<\alpha<\alpha_c$. If
the density asymmetry is smaller than $0.93\alpha_c$, the LOFF
solution will not locate in the conventional LOFF window. In a
recent study of two flavor color superconductivity\cite{gorbar2},
it is found that the LOFF solution satisfying electric charge
neutrality is located at the LOFF1 branch.

We have shown that the LOFF state is thermodynamically stable in
the whole region $0<\alpha<\alpha_c$, where the gap can vary in
the region $0<\Delta<\Delta_0$ and the pair momentum can vary in
the region $0<v_Fq<1.3\delta\mu$. Now we turn to check whether the
LOFF state is dynamically stable in the whole region. Since the
rotational symmetry $O(3)$ is broken down to $O(2)$, the
superfluid density becomes a tensor $\rho_{ij}$. We decompose
$\rho_{ij}$ into a transverse and a longitudinal part due to the
residue symmetry $O(2)$,
\begin{eqnarray}
\rho_{ij}=\rho_T(\delta_{ij}-\hat{q}_i\hat{q}_j)+\rho_L\hat{q}_i\hat{q}_j
\end{eqnarray}
with $\hat{q}\equiv{\vec q}/|\vec{q}|$. Neglecting the terms
proportional to $q/p_F$ and $(q/p_F)^2$ and replacing
$pq\cos\theta/ m$ by $v_Fq\cos\theta$ under the assumption of
small $q$, we get at zero temperature
\begin{eqnarray}
\rho_T=mn\left[1-\frac{3}{4}\int_{-1}^1d\cos\theta\sin^2\theta\eta_\theta
\frac{\delta_\theta^2\theta(|\delta_\theta|-\Delta)}{\sqrt{\delta_\theta^2-\Delta^2}}\right]&&, \nonumber\\
\rho_L=mn\left[1-\frac{3}{2}\int_{-1}^1d\cos\theta\cos^2\theta\eta_\theta
\frac{\delta_\theta^2\theta(|\delta_\theta|-\Delta)}{\sqrt{\delta_\theta^2-\Delta^2}}\right]&&
\end{eqnarray}
with $\eta_\theta$ defined as
\begin{eqnarray}
\eta_\theta=\frac{p_1^3(\theta)+p_2^3(\theta)}{6\pi^2n}.
\end{eqnarray}
Note that the expression $\rho_{ij}$ is the same as the Meissner
mass tensor for the 8th gluon in two flavor LOFF color
superconductor\cite{giannakis2} except for a pre-factor $mn$, if
we employ the good approximation $\eta_\theta\simeq1$. Using the
numerical result for the LOFF solution, we calculated $\rho_T$ and
$\rho_L$ and found that $\rho_T$ is almost zero, which is
consistent with the analytical result in\cite{giannakis2}. The
longitudinal superfluid density $\rho_L$ scaled by $mn$ is shown
in Fig.\ref{fig13} as a function of the scaled gap parameter, it
is positive in the region $0<\Delta<0.83\Delta_0$ and negative for
$0.83\Delta_0<\Delta<\Delta_0$. Hence, the LOFF state with
$0.83<\Delta/\Delta_0<1$, i.e., in the small asymmetry region
$0<\alpha<0.2\alpha_c$ or small LOFF momentum region
$0<v_Fq/\delta\mu<0.4$, is dynamically unstable. For the BP state,
the superfluid density is absolutely negative due to the fact that
the manifold of the gapless excitations in momentum space is very
large. This negative superfluid density is generally cured in the
LOFF state since the manifold of gapless excitations is suppressed
by anisotropy. However, in the region $0.83<\Delta/\Delta_0<1$,
the scaled LOFF momentum $v_Fq/\delta\mu$ which reflects the
degree of anisotropy becomes very small, the superfluid density
will still be negative. When $\Delta/\Delta_0\rightarrow 1$,
$v_Fq/\delta\mu\rightarrow 0$, the anisotropy disappears, and the
superfluid density becomes divergent.

\section {Summary}
\label{s6}
The key theoretic problem in the study of asymmetric fermion
superfluid is the pairing mechanism. The familiar mechanisms
include isotropic BCS and BP states and anisotropic LOFF state.
The question is which of them is the real ground state for a
system under some physical condition which can lead to mismatched
Fermi surfaces of the two fermions.

We have investigated the competition between the normal, BCS, BP
and LOFF phases in asymmetric fermion superfluids in the frame of
a general four fermion interaction model in weak coupling region.
The masses of the two species $m_a$ and $m_b$ are set to be equal,
but their chemical potentials $\mu_a$ and $\mu_b$ can be
different. Two constraints on the system are discussed: (1) fixed
chemical potential asymmetry and (2) fixed fermion number
asymmetry.

In BP phase, there are two kinds of instabilities, one is the
Sarma instability which was found many years ago, and the other is
the magnetic instability which was recently pointed out in the
study of interior gap superfluidity and gapless color
superconductivity. While the magnetic instability gives us a
strong hint that the LOFF phase may be stabler than the BP phase,
to confirm this statement needs the comparison of the free
energies for the normal, BCS, BP and LOFF states, when the
chemical potentials of the two species are not directly fixed. By
solving the gap equations, we obtained all possible solutions, and
then by comparing the thermodynamic potentials $\Omega$ or free
energies $F$ for the four phases, we determined the real ground
state which corresponds to the lowest minimum of $\Omega$ or $F$.
For case (1), the BP state is always unstable due to the Sarma
instability, the system is in BCS phase when the chemical
potential difference is small and in normal phase when the
difference is large enough, and the LOFF state can exist only in a
narrow window of asymmetry. For case (2), the Sarma instability of
the BP state can be avoided due to the disappearance of the BCS
phase, if $n_a$ and $n_b$ are set to be unequal. However, due to
the magnetic instability of the BP state, in the coexistence
region of BP and LOFF, the LOFF state is always the stable one. In
addition, the LOFF state can even survive when the BP disappears
in the highly asymmetric region. We emphasis that the LOFF windows
in case (1) and (2) are quite different. In case (1), the LOFF
state is thermodynamically stable only in a narrow window, i.e.,
$0.707\Delta_0<\delta\mu<0.754\Delta_0$ in weak coupling. In this
window, the gap $\Delta$ can vary in the region
$0<\Delta<0.24\Delta_0$ and the LOFF momentum is almost a
constant, $1.2<v_Fq/\delta\mu<1.3$. However, in case (2), the LOFF
state is thermodynamically stable in a large region
$0<\alpha<\alpha_c$, which corresponds to
$0.635\Delta_0<\delta\mu<\Delta_0$, $0<\Delta<\Delta_0$ and
$0<v_Fq<1.3\delta\mu$. Only in the small $\alpha$ region
corresponding to $0.83\Delta_0<\Delta<\Delta_0$ the superfluid
density is negative, which means that the LOFF state in this
region is dynamically unstable.

We considered in this paper only the asymmetric systems with equal
masses of the two kinds of fermions. However, a system with mass
difference is of great interest, since it can be realized in the
atomic fermion gas with a mixture of different fermionic atoms
such as $^6Li$ and $^{40}K$\cite{liu}. As we have shown in
\cite{he}, the breached pairing state can be free from magnetic
instability when the mass ratio of the two species becomes very
large. The true ground state of these systems is still unknown and
needs further investigation. The superfluidity we considered here
is non-relativistic, which can be applied to the study of
electronic system, atomic fermion gas, and nuclear matter, and can
be extended to relativistic fermion superfluids if the broken
symmetry is Abelian. In the study of color superconductivity where
the broken color symmetry is non-Abelian, the problem of
(chromo)magnetic instability is more complicated. Some approximate
calculations in this direction are presented
recently\cite{giannakis3,gorbar2}.

Nevertheless, the mixed phase or phase separation
\cite{bedaque,caldas} which has been observed in recent
experiments\cite{exp1,exp2} is not included in our discussion. We
have planned to give a more complete work including the important
mixed phase\cite{hep}.

{\bf Acknowledgement:} We thank M.Huang for helpful discussions
and M.Alford, K.Rajagopal and H.Ren for the encouragement during
the work. The work was supported by the grants NSFC10428510,
10435080, 10447122, 10575058 and SRFDP20040003103.

\end{document}